\documentclass[prd,showpacs,preprintnumbers,preprint]{revtex4}
\usepackage{amssymb}
\usepackage{graphicx}
\usepackage{dcolumn}
\usepackage{bm}
\usepackage{amsfonts}
\newcommand{\sial}{\sin\alpha}

\newcommand{\coal}{\cos\alpha}

\newcommand{\var}{\varepsilon}

\newcommand{\la}{\lambda}

\newcommand{\om}{\omega}
\newcommand{\al}{\alpha}

\newcommand{\pa}{{\partial}}
\newcommand{\ga}{\gamma}
\newcommand{\si}{\sigma}

\newcommand{\fr}{\frac}
\newcommand{\de}{\delta}
\newcommand{\lb}{\label}
\newcommand{\ov}{\overline}

\newcommand{\e}{{\rm e}}

\newcommand{\f}{{f}}

\newcommand{\be}{\begin{equation}}
\newcommand{\ee}{\end{equation}}
\newcommand{\ba}{\begin{eqnarray}}
\newcommand{\ea}{\end{eqnarray}}
\newcommand{\non}{\nonumber}
\newcommand{\bw}{\begin{widetext}}
\newcommand{\ew}{\end{widetext}}

\def\d{\partial}
\def\e{{\rm e}}
\begin{document}
\preprint{ESI 1392}
\preprint{hep-th/0310718}
\title{Axion  bremsstrahlung from collisions of global strings}
\author{D.V. GAL'TSOV$^*$,  E.Yu. MELKUMOVA$^*$ and R. KERNER$^\dag$}
\address{$^*$Department of Physics, Moscow State University,119899,
Moscow, Russia\\$^\dag$Universit\`e Pierre et Marie Curie, 4,
place Jussieu, Paris, 75252, France}

\pacs{11.27.+d, 98.80.Cq, 98.80.-k, 95.30.Sf}
\date{\today}

\begin{abstract}
We calculate axion radiation emitted in the collision of two
straight global strings. The strings are supposed to be in the
unexcited ground state,  to be inclined with respect to each
other, and to move in  parallel planes. Radiation arises when the
point of minimal separation between the strings  moves faster than
light. This effect exhibits  a typical Cerenkov nature.
Surprisingly, it allows an alternative interpretation as
bremsstrahlung under  a collision of point charges in $2+1$
electrodynamics. This can be demonstrated by suitable world-sheet
reparameterizations and dimensional reduction. Cosmological
estimates show that our mechanism  generates axion production
comparable with that from the oscillating string loops and may
lead to further restrictions on  the axion window.
\end{abstract}
\maketitle

\section{Introduction}
For more than twenty years  the Peccei-Quinn
axion~\cite{PQ77,W78,Wi78} remains one of the most  serious dark
matter candidates. Its properties depend  essentially on one
unknown parameter, the vacuum expectation value $\f$ marking the
energy scale of the U(1) symmetry breaking. The axion acquires a
small mass  \be m_a\simeq 6\,{\rm \mu eV}\,\frac{10^{12} {\rm
GeV}}{\f}\ee in the QCD phase transition. The quantity $\f$ is a
free parameter of the theory, but it is severely constrained both
by the accelerator physics and the astrophysics, the upper bound
from the combined data being several meV. The lower bound is of
cosmological origin and follows from the requirement that axions
produced during the cosmological evolution do not overclose the
Universe~\cite{AS83}. The cosmological axion production is mainly
due to the axion radiation from global strings and domain walls.
 In view of still persisting theoretical uncertainty, the
corresponding lower bound lies between several units  and several
tens of $\mu$eV, corresponding to the maximal value of $\f$
between $10^{12}$ and $10^{11}$ GeV. (For  recent review of the
present theoretical status of the axion see~\cite{Sr02};  the
current astrophysical status of  this model is reviewed
in~\cite{Si02}, earlier reviews include~\cite{Re}).

The axion string
network~\cite{EV81,Vi85,GHV90,ViSh94,HiKi95,BaSh97,BaSh98,K98,BaSh99}
is formed at the temperature of the Peccei-Quinn phase transition
$\f$, and it is usually assumed that the reheating temperature is
higher than $\f$, otherwise the  network would  get diluted by the
exponential expansion; this introduces more uncertainty into
theoretical predictions. Strings are primarily produced as long
straight segments with length of the order of the horizon size,
and they initially move with substantial
friction~\cite{DS87,VI91,MaSh97} due to scattering on them of the
cosmic plasma particles. At some temperature $T_*<\f$ the
scattering becomes negligible and the string network enters the
scale-invariant regime~\cite{HS91,Na97,YKY98} when strings form
closed loops via interconnections and move almost freely with
relativistic velocities~\cite{AlTu,Sh87}. The standard estimates
of the axion radiation from global strings  are based on the
assumption that the main contribution comes from the oscillating
string
loops~\cite{D85,GaVa,DS89,DaQu90,Sa,BS94,YKY98,HCS99,HaChSi01}.
The amount of radiation is large enough and leaves a rather narrow
window for possible values of $\f$.

Here we  would like to discuss one additional mechanism of axion
radiation: the bremsstrahlung which can be produced in collisions
of long strings. In fact, the number of collisions of long strings
is at least not smaller than the number of string loops which are
formed under some of these collisions (note that in our case
passing of strings through each other is not assumed). Meanwhile,
 as far as we are aware, this mechanism was not explored so far
in the axion physics context. It works as follows. Neglecting the
boundary effects, let us assume that two infinite straight strings
move in  parallel planes,  the two strings being generically
inclined with respect to each other. Then the point of  minimal
separation between the strings, which marks  the localization of
the effective source of  axion radiation, is allowed to move with
any velocity,  superluminal in particular. In this latter case one
can expect Cerenkov  axion emission to emerge. Similar mechanism
was earlier suggested for gravitational radiation of local strings
\cite{GaGrLe93}, but in that case the explicit calculations have
led to zero result. Vanishing of the gravitational radiation,
however, has a specific origin related to the absence of gravitons
in the $2+1$ gravity theory. Indeed, as it was explained in
\cite{GaGrLe93}, two crossed superluminal strings can be
``parallelized'' by suitable coordinate transformations and world
sheet reparameterizations, so the problem is essentially
equivalent to the point  particles' collision in $2+1$ gravity
theory. For other fields, (e.g. electromagnetic) this mechanism
does work; see~\cite{GGL} for calculation of the Cerenkov
electromagnetic radiation generated in collisions of
superconducting strings. So one can expect to have a non-vanishing
axion bremsstrahlung from collisions of global strings as well.

Our calculation follows the approach of~\cite{GaGrLe93} and it is
essentially perturbative in terms of the string-axion field
interaction constant (equal to $\f$) involving two subsequent
iterations in the strings equations of motion and the axion field
equations. This is similar to calculation of the bremsstrahlung
from charged ultrarelativistic particles. In this latter case the
iteration sequence converges if the scattering angle is small. We
assume the same condition to hold for the strings, though we do
not enter here into a detailed investigation of the convergence
problem. We would like to mention also a similar perturbative
approach in terms of geodesic deviations~\cite{KHC}.
\section{General setting}
Consider a pair of relativistic strings
$$x^{\mu}=x_n^{\mu}(\sigma^a),\;\; \mu=0,1,2,3,\;\; \sigma_a=(\tau,
\sigma),\;\; a=0,1,$$ where  $n=1,2$ is the  index labelling the
two strings. The 4-dimensional space-time is assumed to be flat
and the signature is chosen as $+,---$ and $(+,-)$ for the string
world-sheets. Strings interact via the axion field $B_{\mu\nu}$ as
described by the action~\cite{BaSh97} \be\label{fac}
  S=S_{st}+S_B,
\ee  where the string term is \be \label{mac}
S_{st}=-\sum_{n=1,2}\int \Big( \frac{\mu_n^0}{2} \sqrt{-\gamma}
\gamma^{ab}\partial_a x_n^ \mu
\partial_b x_n^ \nu  \eta_{\mu \nu}+
2\pi
 f_{n}  B_{\mu \nu} \epsilon^{ab}
\partial_a x_n^\mu
\partial_b x_n^\nu \Big)d^2 \sigma_n,
\ee and the  field action is
 \be\label{fgac}
S_B= \frac{1}{6 }\int\limits H_{ \mu \nu \la} H^{ \mu \nu \la}
d^4x. \ee Here $\mu_n^0$ are the (bare) string tension parameters,
$\f_n$ are  the  corresponding coupling parameters (their
$n$-labelling helps to control the perturbation expansions), the
 Levi-Civit\`a symbol is chosen as $\epsilon^{01}=1,\; \ga_{ab}$
is the metric on the world-sheets.

The totally antisymmetric axion field strength is defined as \be
H_{ \mu \nu \la}=\partial_\mu B_{\nu\la}+\partial_\nu
B_{\la\mu}+\partial_\la B_{\mu\nu}.\ee Variation of the action
(\ref{fac}) over $ x^\mu_n $ leads to the  equations of motion for
strings \ba\label{em} \mu_n^0
\partial_a(\sqrt{-\ga}\ga^{ab}\partial_b x^\mu_n)=2\pi f_n
H^\mu_{\al\beta} V_n^{\al\beta}, \ea where \be\lb{V} V_n^{\al
\beta} = \epsilon^{ab}
\partial_a x_n^\al
\partial_b x_n^\beta,\ee  and $ H^\mu_{\al\beta} $ is
the total field strength due to both strings (no external field is
assumed). The corresponding potential two-form is the sum \be
B^{\mu\nu}=B^{\mu\nu}_1+B^{\mu\nu}_2,\ee of contributions
$B^{\mu\nu}_n$ due to each string: \be\lb{eeq}
 \square B_n^{\mu\nu}=4\pi J_n^{\mu\nu}.
\ee Here the Lorentz gauge $\pa_\la B^{\nu\la}=0$ is assumed, the
4-dimensional D'Alembert operator  is introduced as
$\Box=-\eta^{\mu\nu}\pa_\mu\pa_\nu$, and the source term reads
\be\lb{mst} J_n^{\mu\nu}= \fr{f_n}{2} \int\limits V_n^{\mu \nu}\,
\delta^4(x-x_n(\si_n))\,d^2 \sigma_n. \ee

The {\em self-action} terms in the equations of motion (\ref{em})
diverge both near the strings and at large distances, so two
regularization parameters $\delta$ and $\Delta$ have to be
introduced~\cite{BaSh95}.  These can be absorbed by  classical
renormalization of the string tension as follows \be\lb{mure}
\mu=\mu_0+2\pi f^2\log (\Delta/\delta).\ee The ultraviolet cutoff
length $\delta$ is of the order of  string thickness $\delta\sim
1/f$, while the infrared cutoff distance $\Delta$ can be taken as
the correlation length in the string network.

Assuming that such a renormalization is performed, we are left
with the same equations of motion (\ref{em}) with the physical
tension parameters $\mu_n$ which contain only the {\em mutual}
interaction terms  on the right hand side. The constraint
equations for each string read : \ba\label{con}
\partial_a x^\mu \partial_b x^\nu \eta_{\mu\nu}- \fr12 \ga_{ab}\ga^{cd}
(\partial_c x^\mu \partial_d x^\nu \eta_{\mu\nu})=0.\ea Using the
invariance of the action (\ref{mac}) under separate world-sheet
reparameterizations \be \si^a_n \; \to \; \ov{\si}^a_n (\si^a_n)
\ee and under Weyl rescaling of the metric $\ga_{ab},$ one can fix
the  {\it flat gauge} \be\lb{gau}\ga_{ab}=\eta_{ab},\ee so that
the constraint equations
 simplify to \be\lb{cons}
 \dot{x}^2 + x'^2  =  0,  \quad    \dot{x}^\mu x'^\nu \eta_{\mu \nu}  =
0 \ee for each string, where dots and primes denote the
derivatives over $\tau$ and $\si$ as usual. In the gauge
(\ref{gau}) the renormalized   equations of motion read : \be
\lb{steq} \mu_n\left(\pa_\tau^2-\pa_\si^2\right)x^\mu_n=2\pi f_n
H^{\mu\nu\la}_{n'}V_{\nu\la}^{n'},\quad n\neq n',\ee where the
field $H^{\mu\nu\la}_{n'}$ corresponds to  the contribution of the
$n'-$th string (no sum over $n'$). It is worth noting that
imposing  flat world-sheet metric we still do not fix the gauge
completely,  since one is free to perform two-dimensional linear
transformations preserving the Minkowski metric (\ref{gau}) on the
world sheets.

The retarded solution to the wave equation can be presented in its
standard form \be\lb{Bret}
   B_n^{\mu\nu}=4 \pi \int G^{ret}(x-x') J_n^{\mu\nu}(x')d^4x',
\ee where the Green's function is \be
    G^{\rm ret}(x-x')
  = \frac{1}{(2 \pi)^4} \int \frac{\e^{-iq(x-x')}}{q^2+2i\epsilon
q^0}d^4q, \ee with  $ \epsilon \rightarrow +0$. The advanced
Green's function is obtained  by changing the sign of $\epsilon$.

The energy momentum tensor for the axion field reads \be
T^{\mu\nu}=\left( H^{\mu\al\beta}H^\nu_{\al\beta} \ - \ \fr{1}{6}
H_{\al\beta\ga}H^{\al\beta\ga}g^{\mu\nu}\right).\ee To avoid
complications with definition of the wave zone for infinite
strings, it is convenient to compute the radiation reaction power
using the divergence equation \be\label{sk} \fr{\partial
T^{\mu\nu}}{\partial x^\nu}  = -4 \pi
H^{\mu\al\beta}J_{\al\beta},\ee where $ H^{\mu\al\beta}$ should be
taken as the half-difference of the retarded and advanced solution
of the wave equation,  $ H_{\rm rad}^{\mu\al\beta}=( H_{\rm
adv}^{\mu\al\beta}- H_{\rm adv}^{\mu\al\beta})/2$. The
corresponding potential  has then the following Fourier transform:
\be\label{br} B^{\rm rad}_{\mu\nu}(k)  =-
4i\pi^2\fr{k^0}{|k^0|}J_{\mu\nu}(k)  \delta (k^2) \ ,\ee where the
current \be J_{\mu\nu}(k)=\int J_{\mu\nu}(x)\e^{ikx} d^4x\ee is
transverse \be\lb{jt} k^\mu J_{\mu\nu}(k)=0\ee and  taken on the
mass-shell $k^2=0$ of the massless axion. The radiative loss of
the 4-momentum during the collision can be  evaluated as
\ba\label{bk} P^\mu \ = \ \int\partial_\nu T^{\mu\nu} d^4x =-4 \pi
\int H_{\rm rad}^{\mu\al\beta}J_{\al\beta} d^4x\ .\ea Substituting
(\ref{br}) into (\ref{bk})
 we obtain \be\label{tot}
P^\mu \ = \ \fr1{\pi}\int k^\mu \fr{k^0}{|k^0|}|J_{\al\beta}(k)|^2
\delta (k^2)d^4k.\ee

Alternatively, this quantity can be presented as a square of the
polarization projection
  of the  Fourier-transform of the current. Indeed, in three
dimensions the axion  field propagating along the wave vector
${\bf k},\;k^\mu=(\omega, {\bf k}),$ has a unique polarization
state \be\lb{pol} e_{ij}=\fr1{\sqrt2}(e_i^\theta
e_j^\varphi-e_i^\varphi e_j^\theta),\quad i, j=1,2,3,\ee where
${\bf e}^\theta$ and ${\bf e}^\varphi$ are two unit vectors
orthogonal to ${\bf k}$ and to each other: \be {\bf
e}^\varphi\cdot {\bf e}^\theta=0,\quad {\bf k}\cdot {\bf
e}_i=0.\ee Using antisymmetry and transversality of the current
(\ref{jt}) and the completeness condition \be e_i^\theta
e_j^\theta + e_i^\varphi e_j^\varphi =\delta_{ij}-k_i
k_j/\omega^2,\ee one finds \be\label{totpol} \Delta P^\mu =
\fr1{\pi}\int k^\mu \fr{k^0}{|k^0|}|J^{ij}(k)e_{ij}|^2 \delta
(k^2)d^4k.\ee Integrating over $k^0$, we finally obtain
\be\label{totpol3} \Delta P^\mu  = \fr1{\pi}\int \fr{k^\mu}{|{\bf
k}|}|J({\bf k})|^2 d^3k, \ee where \be \lb{jka}J({\bf
k})=J^{ij}(k)e_{ij}\ee with $k^0=|{\bf k}|$. In what follows we
shall use the following explicit parameterization of three
orthogonal vectors by  spherical angles :
 \ba {\bf k}&=&
\omega[\sin\theta \cos\varphi , \, \,  \sin\theta \sin\varphi, \,
\,
\cos\theta],\non\\
 {\bf e}^\varphi &=&[-\sin\varphi, \, \, \cos\varphi, \, \, 0],\\
{\bf e}^\theta &=& [\cos\theta\cos\varphi, \, \,
\cos\theta\sin\varphi, \, \, -\sin\theta].\non \ea

Our strategy consists in solving the system of equations
(\ref{steq}, \ref{Bret}) iteratively, by expanding  all functions
in  powers of the coupling constant $\f$: \ba\label{bs} B^{\mu
\nu} &=&   \sum_{l=1}^{\infty} {_lB}^{\mu\nu}, \\ \label{js}
J^{\mu \nu}   &=&   \sum_{l=0}^{\infty} {_lJ}^{\mu\nu}, \\
\label{xs}
   x^{\mu}(\sigma)  &=&
   \sum_{l=0}^{\infty} {_lx^\mu}(\sigma).
\ea Such expansions have to be  performed separately for each
string. In our notation, the  zeroth-order approximations for the
string embedding functions $^{}_0x^\mu_n(\sigma)$ give
zeroth-order source terms $^{}_0J^{\mu\nu}_n$, which generate the
first order fields $^{}_1B^{\mu\nu}_n$  originating on each
string. These quantities,  when substituted to the string
equations of motion~(\ref{steq}), will generate the first order
terms $^{}_1x^\mu_n$, and so forth. Convergence of this expansion
(with a formal parameter $\f$ which has a dimension of an inverse
length) is difficult to explore in general. This is likely to
depend substantially on the choice of the zeroth-order solution of
the string equations of motion. In this paper, the strings are
assumed to be in the unexcited state in the zeroth order, that is
they are straight and freely moving. In this case it can be
expected that the power series solutions will give sensible
results (at least in the lowest appropriate order when radiation
appears) if the deviation angle of colliding strings due to their
lowest order interaction is small.   We shall come back to this
condition later on.
\section{  Collision kinematics}
\begin{figure}
\centering
\includegraphics[width=6cm,height=6cm,angle=-90]{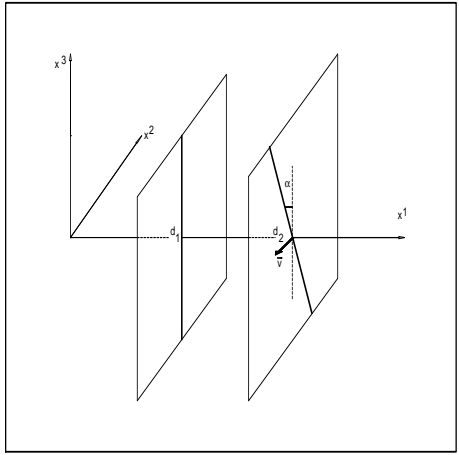}
\caption{In the chosen frame, the first string is at rest, the
second  moves in the parallel plane separated from the first by an
impact distance $d=d_2-d_1$ being inclined at the angle $\al$ with
respect to the string at rest. Its velocity $v$ is perpendicular
to the string itself.}
\end{figure}
The kinematics of  collision is shown on Fig. 1. The unperturbed
world-sheets of unexcited, freely moving straight (infinite)
strings are described by linear functions of $ \tau,\, \sigma $
:\be\label{xo}
   _0x^\mu_n \ = \ d^\mu_n \ + \ u^\mu_n \tau_n \
+ \Sigma^\mu_n \sigma_n , \ \ \ n=1,2, \ee
 with  $ d^\mu_n  $ and $ \Sigma ^\mu_n $  being space-like
and $ \ u^\mu_n  $ time-like constant vectors. Then the constraint
equations (\ref{cons}) imply the orthogonality and normalization
conditions (choosing the unit norm time-like 4-velocity $u^\mu$ )
\be\label{usig} u^\mu \Sigma_\mu=0, \quad u^\mu u_\mu\
=1=-\Sigma^\mu \Sigma_\mu \ee for each string. We choose the
reference frame such that the first string is at rest and is
stretched along the $x^3$ axis: \be u^\mu_1=[1,0,0,0],\quad
\Sigma^\mu_1=[0,0,0,1].\ee The second string is assumed to move in
the plane $x^2,\,x^3$ with the velocity  $ v $ orthogonal to the
string itself: \be u^\mu_2=\ga \, [1,0,-v\coal,v\sial],\quad
\Sigma^\mu_2=[0,0,\sial,\coal],\ee where $\ga  = (1-v^2)^{- 1/2}
$. We also choose  both impact parameters $d^\mu_n$ to be
orthogonal to $u^\mu$ and $\Sigma^\mu$ and aligned with the axis
$x^1$, the distance between the planes being $d=d_2-d_1$. The
angle of inclination $\al$ of the second string with respect to
the first one can be presented in the Lorentz-invariant form as
\be\label{ang} \alpha = \arccos(- \Sigma_1 \Sigma_2), \ee and the
relative velocity of the strings as \be \lb{vel} v \ = \
(1-(u_1u_2)^{-2})^{ \frac {1}{2}}. \ee With this parametrization
of the unperturbed world-sheets, the point of  minimal separation
between the strings (which always remains  equal to $d$) moves
with the velocity \be\lb{vp} v_p= \frac{v}{\sin \alpha}=(u_1
u_2)^{-1}\left( \frac{(u_1u_2)^2-1}
  {1-( \Sigma_1 \Sigma_2)^2}\right)^{ \frac{1}{2}}
\ee along the $x^3$-axis. This motion is not associated with
propagation  of signal of any kind, so the velocity $v_p$ may be
arbitrary, in particular, superluminal. The case of parallel
strings corresponds to $v_p=\infty$.

Let us explore the residual gauge invariance consisting of the
SO(1,1) transformations of the world-sheets \ba\lb{O11} \tau_n
&\to& \tau_n \cosh \chi_n+\sigma_n\sinh\chi_n, \non\\ \sigma_n
&\to& \tau_n \sinh\chi_n+\sigma\cosh\chi_n,\ea preserving the
constraints (\ref{cons}). It is easy to show that the relative
orientation of two strings is not a gauge independent
property~\cite{GaGrLe93}. One can build the following matrix
describing the relative orientation of the world-sheets \be
\kappa_{ab}=\d_a x_1^\mu \d_b x_2^\nu\eta_{\mu\nu},\ee whose
determinant \be \kappa=\det\kappa_{ab}=\gamma\cos\al\ee is
invariant  under the transformation (\ref{O11}). But the relative
velocity (\ref{vel}) and the inclination angle (\ref{ang})
separately are not invariant. In particular, for superluminal
strings one can always find  the world-sheet boosts which make the
strings parallel. This is achieved by the following choice of
parameters in (\ref{O11}) for two strings \be
\tanh\chi_1=\fr{\sin\al}{v},\quad \tanh\chi_2= \fr{\tan\al}{\gamma
v},\ee which exists if $v_p>1$. Note that strings are superluminal
in the above sense provided  that \be \kappa=\ga\cos\al\geq 1.\ee
If this condition is not fulfilled,  then the ``parallelizing''
transformation does not exist.
\section{First order interaction}
According to  the conventions chosen above, the field equations
(\ref{eeq}) for the first order field terms $^{}_1B^{\mu\nu}_n$
read :
 \be\lb{eeqo}
 \Box \, {^{}_1B}_{n}^{\mu\nu }=4\pi\,{_0J}_{n}^{\mu\nu},
\ee with the   source terms (\ref{eeqo})
 \be\lb{jo}  _0{J^{\mu \nu}_{n}}= \fr{f_n}{2} \int\limits
 \,V_{n}^{\mu \nu}
\delta^4 \left(x- {_0x}_n(\sigma_n)\right)\,d^2\sigma_n, \ee where
the zeroth-order quantities are  found under the right-hand side
integral : \be
 V_{n}^{\mu \nu} =u_{n}^\mu \Sigma_{n}^\nu - \Sigma_{n}^\mu
 u_{n}^\nu. \ee
In what follows we pass to the momentum representation according
to the convention : \be F(q)=\int F(x)\,\e^{iqx}\,d^4x, \ee where
$qx=q_\mu x^\mu$.   We will use similar notation for  all scalar
products,  when there is no risk of ambiguity. The Fourier
transform
 of the zeroth-order current (\ref{jo}) then reads :
 \be  ^{}_0J^{\mu \nu}_{n}(q)=2 \pi^2 f_n
\e^{iqd_n}\delta(q u_n)\delta(q\Sigma_n)V_{n}^{\mu\nu}.
 \ee
The retarded solution ${_1B}^{\mu\nu} $ of the Eq. (\ref{eeqo})
generated by the zero-order source is \be\label{b1}
^{}_1B_{n}^{\mu\nu}(q)=-\frac{8\pi^3 f_n V^{\mu\nu}_n
\e^{iqd_n}\delta(q u_n)\delta(q \Sigma_n )}{q^2+2i\epsilon q^0}.
\ee  Note that two $\delta$-functions in this expression  shift
the momentum $q$ from the pole $q^2=0$, so, as can be easily seen
from the Eq. (\ref{tot}), the first order field is non-radiative.

The next step is to find corrections to the world-sheet embedding
functions due to  interaction  between the strings via the axion
field. Substituting (\ref{b1}) into the right hand side of the
equations of motion (\ref{steq}) we obtain the following
first-order perturbations of the string world-sheets
\be\label{x1} ^{}_1x_1^{ \mu} = i \fr{2f _1 f_2}{ \mu_1}\int
X^{\mu}_1 \frac{\delta(qu_2)  \delta(q \Sigma_2) \
\e^{iq(d_2-d_1-u_1\tau-\Sigma_1\sigma)}}
{q^2[(q\Sigma_1)^2-(qu_1)^2]} d^4q, \ee   where \ba X^{\mu}_{1}&=&
q^\mu A + B_{1}\Sigma^\mu_{2} + C_{1}u_2^\mu,\quad
A=(u_1u_2)(\Sigma_1 \Sigma_2)-( \Sigma_1 u_2)(u_1 \Sigma_2),
\non\\  B_1&=&(qu_1)(u_2 \Sigma_1)-(\Sigma_1q)(u_1u_2), \quad
C_1=(u_1 \Sigma_2)( \Sigma_1q)-(qu_1)(\Sigma_1 \Sigma_2)  \ . \ea
Similarly for the second string \be\label{x2} ^{}_1x_2^{ \mu}
=i\fr{2f_1 f_2}{\mu_2}\int X^{\mu}_2 \frac{\delta(qu_1) \ \delta(q
\Sigma_1 ) e^{iq(d_1-d_2-u_2\tau-\Sigma_2\sigma)}}
{q^2[(q\Sigma_2)^2-(qu_2)^2]}  d^4q, \ee  where
 \ba X^{\mu}_{2}= q^\mu A &+& B_{2}\Sigma^\mu_{1} +
C_{2}u_1^\mu,\quad  B_2=(qu_2)(u_1 \Sigma_2)-(\Sigma_2q)(u_1u_2),
\non\\  C_2&=&(u_2 \Sigma_1)( \Sigma_2q)-(qu_2)(\Sigma_1
\Sigma_2),\ea
with the same $A$. It can be checked that the
quantities $X^\mu_n$ satisfy  the conditions \be X^\mu_1
u_{1\mu}=X^\mu_1 \Sigma_{1\mu}=X^\mu_2 u_{2\mu}=X^\mu_2
\Sigma_{2\mu}=0,\ee which ensure the fulfilment of the constraint
equations (\ref{cons})  up to the first order.

To estimate the scattering angle $\delta \varphi$ of the string
collision let us consider the case of parallel strings, $\al=0$.
In the chosen frame one can use as an estimate for $\delta
\varphi$ the ratio of the limiting value of the $x^1$ component of
the 4-velocity of the second string  and the initial $x^2$
component: \be \delta \varphi \sim
\lim_{\tau\to\infty}\fr{^{}_1{\dot x}_2^1}{^{}_0{\dot
x}_2^{2}}.\ee To calculate the quantity in the numerator one has
to integrate in (\ref{x2}) over $d^4q$. Integration over $q^0$ and
$q^3$ is performed using delta-functions. Then, in the limit
$\tau\to\infty$, an asymptotic relation \be
\lim_{\tau\to\infty}\fr{\sin[(qu_2)\tau]}{qu_2}=\pi\delta(qu_2)\ee
gives one more delta-function, allowing  integration over $q^2$.
The last integration over $q^1$ reduces to \be
\int_{-\infty}^{\infty}\frac{\e^{iq^1d} dq^1}{q^1}=i\pi,\ee the
resulting estimate for the scattering angle being \be\delta
\varphi \sim \fr{2\pi^2 f^2}{\mu\,\gamma\, v^2},\ee where we have
set $f_1=f_2$. Using the Eq. (\ref{mure}) for the renormalized
string tension, in which we assume the second term to be
dominant~\cite{BaSh95}, we obtain \be\delta \varphi \sim
\fr{\pi}{\log{(\Delta/\delta)}\,\gamma\, v^2}\,.\ee  The smallness
of this quantity may serve as a rough estimate of the validity of
the perturbation expansions (\ref{bs}-\ref{xs}). Therefore our
iteration procedure works in the relativistic case $v\sim 1$,
especially for ultrarelativistic collisions, $\gamma\gg 1$. But
for global  cosmological strings the quantity
$\log{(\Delta/\delta)}$ is large enough.  This improves  the
convergence of our iteration scheme even for $\gamma\sim 1$.
Therefore our approach seems to be reasonably justified in
realistic cosmological situations.
\section{Radiation}
Using the first order corrections to the world-sheets of the
string due to their lowest order interaction  terms one can
evaluate higher order  terms of the expansion. Radiation emerges
in the second approximation for the axion field $_2B^{\mu\nu}$
generated by the source term $_1J^{\mu\nu}$ which will be obtained
 from  the first order contributions to the world-sheet
embedding functions as follows:
\ba\label{j1} {_1J}^{\mu \nu}
= \sum_{n=1,2} f_n \int\limits
 d^2 \sigma \left[
 \left(  {_0\dot{x}}_{n}^{[\mu}  {_1x'^{\nu]}_n} +
  {_1\dot{x}}_{n}^{[\mu} {_0x'^{\nu]}_n}\right) -  \fr12V_{n}^{\mu
\nu} \,{_1x}_{n}^\lambda\,\partial_\lambda\right] \delta^{4} (x-
{_0x}_n (\sigma_n)),\ea  where  brackets   denote
anti-symmetrization over indices with the factor $1/2$.  This
follows from the fact that the Fourier-transform of this current
is non-zero on the axion mass-shell $k^2=0$.

  The Fourier transform of the
first-order field source   can be presented in the following form:
\be\lb{jtwo}_1J^{\mu \nu}(k)=-8\pi^2 f_1 f_2 \int \delta(qu_2) \
\delta(q \Sigma_2 ) \delta[(k-q)u_1] \ \delta[(k-q)\Sigma_1]
  \e^{i(d_1k)+q(d_2-d_1))} \left( \frac{f_1Q^{\mu\nu}_1}{\mu_1} \ +
  \  \frac{f_2Q^{\mu\nu}_2}{\mu_2}    \right) \ , \ee
  where  the two terms
 \be
Q^{\mu\nu}_1 \ = \fr{(q \Sigma_1) u_1^{[\mu} {X_1^{\nu ]}}- (q
u_1) \Sigma_1^{[\mu} {X_1^{\nu ]}} -\fr12 V_1^{\mu \nu}
(X_{1\lambda} k^\la) }{q^2[(q \Sigma_1)^2-(qu_1)^2]} \ ,\ee
 \be Q^{\mu\nu}_2 \ = \fr{ [(k-q)
\Sigma_2] u_2^{[\mu} {X_2^{\nu ]}}- [(k-q) u_2] \Sigma_2^{ [\mu}
{X_2^{\nu ]}} -\fr12 V_2^{\mu \nu}({X_{2\lambda}} k^\la)}
{(k-q)^2\{[(k-q) \Sigma_2]^2-[(k-q)u_2]^2\}}  \ee
could be
associated with contributions of the first
  and the second  string, respectively. It has to be noted,
  however, that in the second order the axion field is generated
collectively by a source  term containing symmetrically the
contributions of the first order  coming from both strings. It can
be checked that the antisymmetric tensors $Q^{\mu\nu}_n$ satisfy
the transversality conditions \be k_\mu Q^{\mu\nu}_1=k_\mu
{Q^{\mu\nu}_2}=0\ee ensuring  in turn the transversality of the
current $_1J^{\mu \nu}(k)$.

 Integration over $q$-space is performed as follows. First, using
$ \delta[(k-q)u_1] $, we integrate over $ q^0 $,  which fixes the
value \ba q^0=k^0=\om .\ea  Then, using the delta-functions $
\delta(qu_2) $ and $ \delta(q\Sigma_2) $,  one integrates over $
q^2 $ and $ q^3 $ obtaining \be q^2=-\fr{\om\cos\al}{v} \ , \ \
q^3=\fr{\om\sin\al}{v}. \ee The remaining  delta-function $
\delta[(k-q) \Sigma_1] $ does not further fix the vector $q^\mu$,
but  imposes an extra relation on the parameters :
 \ba
k^3=\fr{\om\sin\al}{v}, \ea which is nothing but the Cerenkov
condition for radiation emitted by an effective source moving with
the velocity $ v_p $ along the $x^3$ axis. Clearly, this motion is
the motion of an effective source localized in the region
surrounding the point of minimal separation between the strings
where their deformation described by the first order interaction
is maximal. This condition can be rewritten as the usual
Cerenkov's cone condition on the angle of emission \be
\cos\theta=1/v_p,\ee which makes sense if $v_p\geq 1$. So the
axion radiation is emitted inside the Cerenkov cone oriented along
the $x^3$  axis (Fig. 2). In the limiting case $v_p=1$, radiation
is emitted strictly along the $x^3$ direction; this occurs when
the string inclination angle $\al$ is related to the string
Lorentz factor as~\be \cos\al=\gamma^{-1},  \, \ \,  \ \ {\rm
with} \, \ \ \, \ \, \ \, \gamma = (1 - (u_1u_2)^{-2}
)^{\frac{1}{2}} .\ee For $\cos\al < \gamma^{-1}$ no radiation is
emitted at all. Therefore, string axion bremsstrahlung has
essentially Cerenkov nature.
 But, rather surprisingly, it has another interpretation as
  a point charge bremsstrahlung in $2+1$ electrodynamics. We
will come back to this in Sec. 7.

Integration over $q^1 $ is performed using the contour integration
in the complex plane. Denoting the impact parameter $ d=d_2-d_1 $
we obtain  for the first term or (\ref{jtwo}) \ba\label{I1}
\int\fr{\e^{-iq^1d}dq^1 f_1(q^1)}{(q^1)^2+\kappa_1^2}=\fr{\pi
f_1(-i \kappa_1)\e^{-\kappa_1 d}}{\kappa_1} ,\ea where  $f_1$
stands for the remaining integrand, and \ba \kappa_1=\fr{\om}{\ga
v}.\ea Similarly, the second term   in (\ref{jtwo}) gives the
integrals \be\label{I2} \int\fr{e^{-iq^1d}dq^1
f_2(q^1)}{(k^1-q^1)^2+\kappa_2^2}=\fr{\pi
f_2(-i\kappa_2)e^{-\kappa_2 d+ik^1d}}{\kappa_2} ,\ee where \be
\kappa_2=\fr{\om\xi}{ v},\quad \xi=\cos\al+\fr{k^2}{\om}v. \ee

Therefore, the whole integral over $d^4q$ of the first term in
(\ref{jtwo}) is proportional to the value of the integrand in the
complex point $q^\mu={\bar q}^\mu$ where \be\label{qnew} {\bar
q}^\mu = \fr{\om}{v} \, [ v,\, -i/\ga,\,-\cos\al,\,\sin\al ],\ee
with the corresponding value of $X_1^\mu$ being \ba\label{x1new}
{\bar X}_1^\mu = \fr{\om}{v}\,  [0,\,i\cos\al,\,1/\ga,\, 0]. \ea
For the second term $Q_2^{\mu\nu}$ one finds instead the following
quantities~\be k^\mu-{\bar{\;\bar q^\mu }}=\fr{\om\xi}{v} \, [
0,\,i,\,1,\,0], \ee \be\label{xnew} {\bar{ \;\,\bar X_2^\mu }} =
\fr{\ga\om\xi}{v} \, [v,\,-i\cos\al,\, -\cos\al,\,\sin\al].\ee

Finally, projecting the current on the polarization tensor
(\ref{pol}), we obtain the following lowest order approximation
for the quantity (\ref{jka}):
 \be\label{jf} _1J({\bf
k})=\fr{4\sqrt{2}\pi^3 f_1 f_2}{\om^2\ga v\sin\theta}
\delta(\cos\theta-1/v_p)\cdot\ee
\[ \left(\fr{f_1 \e^{-\om d/v\ga+ikd_1}}{\mu_1}
(\cos\varphi-i\ga\cos\al\sin\varphi)-\fr{f_2 \e^{-\om d \xi/v
+ikd_2}}{\mu_2\ga^2\xi^2}
 \left[\cos\varphi-i (v\xi\ga^2\sin\theta +\sin\varphi)\right]\right),
 \]
 where
$\xi=\cos\al+v\sin\theta\sin\varphi$. Two terms here may be
attributed to  the contributions of the first (remaining at rest)
and the second (moving) strings; in what follows we set
$f_1=f_2=f,\,\mu_1=\mu_2=\mu$. The total amplitude  is therefore a
superposition of two terms exhibiting different frequency cutoff.
The first  term's exponential factor leads to the isotropic
condition \be\lb{om1}\om\lesssim \fr{v\ga}{d},\ee while the second
gives a $\varphi-$dependent condition \be\lb{om2}\om\lesssim
\fr{v}{d\xi}=\fr{v}{d(\cos\al+v\sin\theta\sin\varphi)}.\ee In
other words, the distribution of radiation on the Cerenkov cone is
anisotropic.  This feature becomes especially pronounced in the
ultrarelativistic case.
\begin{figure}
\centering
\includegraphics[width=6cm,height=6cm,angle=-90]{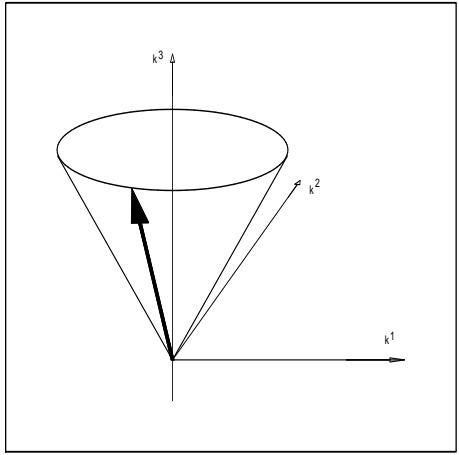}
\caption{Radiation peaking on Cerenkov cone in the direction of
motion of the relativistic sting.}
\end{figure}
Using the identity \be\cos^2\al=v^2\sin^2\theta+\ga^{-2}\ee which
holds on the radiation cone, it is easy to show that when
$\ga\to\infty$, the quantity $\xi$ has a sharp minimum at
$\varphi=-\pi/2$ corresponding to the direction of the moving
string in the rest frame of the first string (see Fig. 2):
\be\lb{xiultra}\xi\thickapprox
\fr{1}{2\ga^2\cos\al}\left(1+\beta^2\ga^2\cos\al^2\right)],
 \ee where
$\beta=\pi/2+\varphi\ll 1$. Due to the factor $\xi^{-2}$ in the
second term in the amplitude (\ref{jka}), radiation is peaked in
the direction $\varphi=-\pi/2$ within the narrow region
\be\lb{bega} \beta\lesssim \ga^{-1}.\ee The exponential factor
exhibits similar peaking, provided \be \ga\cos\al\gg1,\ee so the
frequency range extends up to
\be\om\lesssim\fr{\ga^2}{d\,\cos\al}\ee in the angular region
(\ref{bega}).

At the Cerenkov threshold \be v=\sin\al,\ee the cone of emission
shrinks $\theta\to 0$, in this case \be\xi=\cos\al=\ga^{-1}\ee and
the  amplitude (\ref{jf}) can be considerably simplified. Since in
this limit $kd_n\to 0$, the oscillating factors $\exp{(ikd_n)}$
become equal to unity, and one can check that the leading
contributions of two terms in (\ref{jf}) cancel while $\theta\to
0$, that is the whole expression does not diverge. The remaining
finite contribution reads \be\label{jf0} _1J({\bf
k})=\fr{4i\sqrt{2}\pi^3 f^3}{\mu\om^2}\e^{-\om d/v\ga}
\delta(\cos\theta-1). \ee Inserting this into the Eq.
(\ref{totpol3}) and introducing the normalization length $l$ via
the relation \be
\delta^2(\cos\theta-1)=\frac{l\om}{2\pi}\delta(\cos\theta-1),\ee
one obtains after integration over angles in $d^3k=\om^2 d\om
d\cos\theta d\varphi$  the total radiated energy per unit length
of the string at rest \be \fr{P^0}{l}=\int_{\Delta^{-1}}^\infty
\frac{(2\pi)^5 f^6}{\mu^2\om} \exp({-2\frac{\om d}{v\ga}})d\om,\ee
where $\Delta$ is the frequency cutoff introduced in order to
remove the infrared divergence of the spectral distribution of
radiation. At the high frequency end the spectrum extends till the
boundary (\ref{om1}). Integration over frequencies is performed in
terms of the integral exponential function \be\fr{P^0}{l}=
\frac{(2\pi)^5 f^6}{\mu^2}  {\rm
Ei}\left(1,\frac{2d}{v\ga\Delta}\right).\ee For $d\ll v\ga\Delta$,
\be \fr{P^0}{l} \approx \frac{(2\pi)^5 f^6}{\mu^2}\ln
\left(\frac{v\ga\Delta}{2d\e^C}\right),\ee where $C$ is the Euler
constant, $\e^C=1.781072418$. For $d\gg v\ga\Delta$, one has \be
{\rm Ei} \left(1, \frac{2d}{v\ga\Delta}\right)\approx
\frac{2d}{v\ga\Delta} \exp\left(-\frac{2d}{v\ga\Delta}\right), \ee
in this case radiation is exponentially small. A particular value,
${\rm Ei}(1,1)=0.219,$ can be used to estimate the energy loss for
intermediate impact parameters. For $d= v\ga\Delta$ one has
\be\lb{thre}\fr{P^0}{l}= 2148\,\frac{ f^6}{\mu^2}.\ee

For generic values of parameters $\alpha, \,v$ it is convenient to
rewrite the amplitude in terms of the quantity $\kappa=\ga\cos\al$
which is invariant under world-sheets reparameterizations
characteristic of the relative motion of the strings. Setting
without lack of generality $d_1=0$ we obtain:
\ba\label{jfkappa} _1J({\bf k})=\fr{4\sqrt{2}\pi^3 f^3
\delta(\cos\theta-1/v_p)}{\mu\om^2\sqrt{\kappa^2-1}} \Biggl\{
\exp\left(-\fr{\om
d}{v\ga}\right)\left(\cos\varphi-i\kappa\sin\varphi\right)-\non\\
-\exp\left[-\fr{\om
d}{v\ga}\left(\kappa-i\e^{i\varphi}\sqrt{\kappa^2-1}
 \right)\right]\fr{
 \cos\varphi-i\kappa\left(\kappa\sin\varphi+
\sqrt{\kappa^2-1}\right)}{\left(\kappa+
\sqrt{\kappa^2-1}\sin\varphi\right)^2}\Biggr\}.
 \ea
This is still too complicated to perform  the subsequent
integration analytically, so we pass to the limit
$\kappa\to\infty$, assuming $v\to 1$ and the inclination angle
$\al$ not close to the threshold of  shrinking cone. Then the
second term exhibits peaking around the direction $\varphi=-\pi/2$
which marks the plane of motion of the relativistic second string
(recall that the first string is at rest in the Lorentz frame
chosen).   Keeping only the second term which is dominant in the
main frequency range we obtain the spectral-angular distribution
in the vicinity of this direction as follows \be
\lb{Pkapbom}\fr1{l}\fr{dP^0}{d\om d\beta}=\fr{64\pi^4
f^6\kappa^2}{ \mu^2\om}
\fr{(1-\kappa^2\beta^2)^2}{(1+\kappa^2\beta^2)^4}
\exp\left({-\fr{\om d(1+\kappa^2\beta^2)}{\ga\kappa}}\right).\ee
The spectrum exhibits  an infrared divergence, and in the forward
direction $\beta=0$ it extends up
 to $\om\sim\om_{{\rm max}}$, where \be  \om_{{\rm
 max}}=\fr{\ga\kappa}{d}.\ee Choosing as an infrared cutoff an inverse
length
 parameter $\Delta$ and integrating and over frequencies  we
 obtain the angular distribution of the total radiation
\be\lb{Pkapa} \fr1{l}\fr{dP^0}{ d\beta}=\fr{64\pi^4 f^6\kappa^2}{
\mu^2} \fr{(1-\kappa^2\beta^2)^2}{(1+\kappa^2\beta^2)^4} {\rm  Ei}
\left(1, \fr{d(1+\beta^2\ga\kappa)}{\ga\kappa\Delta}\right) .\ee
Since the integral exponential function decays exponentially for
large argument, the total radiation is peaked around $\beta=0$
within the angle \be\beta\lesssim \sqrt{\ga\kappa}.\ee

\begin{figure}
\centering
\includegraphics[width=6cm,height=6cm,angle=-90]{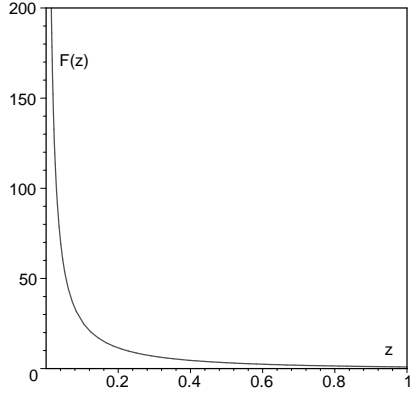}
\caption{The spectral function $F(z)$.}
\end{figure}
One can also obtain the spectral distribution of radiation by
integrating (\ref{Pkapbom}) over the variable $\beta$, which can
be extended to the full axis in view of the exponential decay of
the integrand. The result reads: \be \lb{Pom} \fr1{l}\fr{dP^0}{
d\omega}=\fr{16\pi^5 f^6 d}{3\ga \mu^2}F(z),\quad z=\fr{\omega
d}{\ga\kappa},\ee where \be F=(8z^2+8z+6)\fr{\e^{-z}}{\sqrt{\pi
z}}+(8z^2+12z+6-3/z)({\rm erf}(\sqrt{z})-1).\ee The spectrum is
shown on the Fig. 3. For small frequencies $\omega\ll \ga\kappa/d$
the function $F(z)$ has a logarithmic divergence \be F(z)\sim
\fr3{z},\ee while for large $z$ it decays exponentially \be
F(z)\sim \fr{12\e^{-z}}{\sqrt{\pi z^3}}.\ee Finally, the total
radiation per unit length is obtained by integrating  (\ref{Pom})
over frequencies using a cutoff length
$\Delta$:\be\lb{final}\fr1{l} P^0 =\fr{16\pi^5 f^6\kappa}{3
\mu^2}\phi(y),\quad y=\fr{d}{\ga\kappa\Delta},\ee where\be
\phi(y)=12\sqrt{\fr{y}{\pi}}\,
{_2F_2}\left(\fr12,\fr12;\fr32,\fr32;-y\right)
-3\ln\left(4y\e^{C}\right)+\fr72,\ee where the generalized
hypergeometric function is used. The function $\phi(y)$ is plotted
on Fig. 4, for small $y$ the leading term is the logarithm, while
the asymptotic value for large $y$ is $\phi(\infty)=7/2$.

\begin{figure}
\centering
\includegraphics[width=6cm,height=6cm,angle=-90]{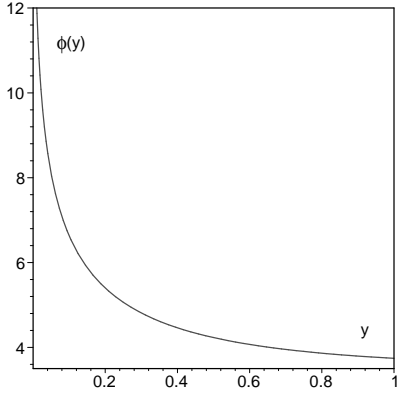}
\caption{Dependence of radiation energy on the impact parameter,
$y={d}/({\ga\kappa\Delta})$.}
\end{figure}

The speed of the effective radiation source $v_p$ tends to
infinity when  the two strings become parallel. In this case
(corresponding to $\kappa=\ga$ in the above formula) the Cerenkov
cone opens up to $\theta=\pi/2$, and the whole picture becomes
essentially $2+1$ dimensional. Since this case is in fact generic
(superluminal strings can be always ``parallelized'' by the
world-sheet reparameterizations (\ref{O11}) and suitable
space-time Lorentz transformations), this gives an alternative
description of the effect as bremsstrahlung under collision of
charges in $2+1$ electrodynamics.
\section{Bremsstrahlung in 2+1 electrodynamics}
Classical system of parallel strings interacting via the
antisymmetric two-form potential $B_{\mu\nu}$ can be equivalently
presented as the system of point charges interacting via the
Maxwell field $A_\mu $ in 2+1 dimensions (in what follows in this
section the Greek indices run over $\mu=0,1,2$). Indeed,
substituting the constraint into the action (\ref{fac}) and
assuming that all variables are $x^3$-independent, one can rewrite
the action  as \be \label{macc} S= -\sum_{n=1,2} \int\limits
\left(m_n d\tau_n + e_n A_{\mu}dx_n^\mu\right)-\fr14\int\limits
F_{\mu\nu}F^{\mu\nu}d^3x.\ee The correspondence between the
quantities here and in the previous section looks as follows:
\be\lb{cors} A_\mu=\sqrt{2l}B_{3\mu},\quad m_n=\mu_n l,\quad
e_n=2\sqrt{2l}\pi f_n,\ee and $F_{\mu\nu}=\partial_\mu
A_\nu-\partial_\nu A_\mu$, where $l$ is the normalization length
along the suppressed coordinate $x^3$. The  equations of motion in
the gauge $\d_\mu A^\mu=0$ read \ba m_n\ddot x^\mu_n &=& e_n
{F^\mu}_\nu\dot x^\nu_n,\\ \square A^\mu &= &-J^\mu,\ea where
$J^\mu=J_1^\mu+J_2^\mu$, \be J^\mu_n=e_n \int\limits\dot
x_n^\mu\de^3(x-x_n(\tau_n))d\tau_n,\ee and only mutual interaction
 should be taken in the particle equations of motion after
the classical mass renormalization.

Performing standard calculations we obtain for the energy loss
under particle collision:\be P^\mu=\fr1{8\pi^2}\int\limits
k^\mu\fr{k^0}{|k^0|}\de(k^2)|J_\mu(k)|^2d^3k.\ee In $2+1$
dimensions a photon has the unique polarization state, so
alternatively this expression can be rewritten (after integration
over $k^0$) as \be \lb{P1} P^\mu=\fr1{8\pi^2}\int\limits
\fr{k^\mu}{|{\bf k}|}|J({\bf k})|^2d^2k,\ee where we put
$k^0=|{\bf k}|,\; e^\mu=(0,{\bf e}),\; J({\bf k}) = J_\mu({\bf k})
e^\mu $, and two-dimensional vectors are parameterized as \ba {\bf
k} &=& \om \,
[\cos\varphi, \sin\varphi],\non\\
{\bf e} &=& [-\sin\varphi, \cos\varphi].\ea

Perturbation theory with respect to the formal charge parameter
$e$ (whose  dimension in the $2+1$ electrodynamics is $l^{-1/2}$
in the units $\hbar=c=1$) is constructed in the same way as in the
Sec. II, introducing the input world-lines \be\label{xoo}
   ^{}_0x^{\mu} _{n} = d^\mu_{n} + u^\mu _{n} \tau_n,
\ee with space-like impact vectors $d^\mu_{n}$ and unit time-like
3-velocities $u^\mu _{n}$: \be\label{uaa} u^\mu_1 = [1,0,0], \quad
u^ \mu_2 =  \gamma [1,0,-v]. \ee After the first iteration one
obtains \be ^{}_1x^{\mu}_{1} =\fr{ie_1e_2}{4\pi^2m_1}\int\limits
\fr{u_2^\mu(qu_1)-q^\mu(u_1u_2)}{q^2(qu_1)^2}\de(qu_2)\e^{iq(d-u_1\tau_1)}d^3q,\ee
where $d=d_2-d_1$, and similarly for the second charge. The first
order current,   which gives the lowest order contribution to
radiation, reads \be _1J^\mu=\sum_{n=1,2} e_n\int\limits
d\tau_n\left(^{}_1{\dot x}^\mu_n-{u^\mu_n}\,
^{}_1x^\nu_n\d_\nu\right)\de\left(x- {_0x_n}(\tau_n)\right).\ee
Its Fourier-transform is
\be\lb{jtwop}_1J^{\mu}(k)=\fr{e_1
e_2}{2\pi} \int \delta(qu_2) \delta[(k-q)u_1]
  \e^{i(d_1k+qd)} \left( \frac{e_1}{m_1}Q^{\mu}_1  +
 \frac{e_2}{m_2} Q^{\mu}_2   \right)d^3 q, \ee
  where
 \ba
Q^{\mu }_1 &=& \fr{(ku_1)^2
u_2^\mu-(ku_1)(ku_2)u_1^\mu+(u_1u_2)[(kq)u_1^\mu-(ku_1)
q^\mu]}{q^2(ku_1)^2},\non\\
 Q^{\mu }_2 &=& \fr{(ku_2)^2 u_1^\mu-(ku_1)(ku_2)u_2^\mu+(u_1u_2)
 [(kq)u_2^\mu-(ku_2)(k-q)^\mu]}{(k-q)^2(ku_2)^2}.
\ea
Integration over $q$ is performed similarly  as in the
case of strings: two delta-functions are used to integrate over
$q^0,\,q^2$, the remaining integral over $q^1$ is evaluated using
a contour integration in the complex plane. Finally, projecting
the amplitude on the polarization vector $e^\mu$ we obtain
\be\label{jfp} _1J({\bf k})=\fr{ e_1 e_2}{2\om \ga v}
\left(\fr{e_1\e^{-\om d/v\ga+ikd_1}}{m_1}
(-\cos\varphi+i\ga\sin\varphi)-\fr{e_2 \e^{-\om d \xi/v
+ikd_2}}{m_2\ga^2\xi^2}
 \left[\cos\varphi-i (v\xi\ga^2 +\sin\varphi)\right]\right),
 \ee
 where $\xi=1+v\sin\varphi$.

As we argued in the Sec. 3, our iteration scheme converges if the
relative velocity is relativistic, and for cosmological
applications one is interested by moderately relativistic
velocities when $\ga$ is not large.  However, for general $\ga$ it
is difficult to perform computation of the total energy loss
analytically, so we continue calculations in the ultrarelativistic
case $\ga\gg 1$ when additional simplifications can be made.
Hopefully this should give a reasonable approximation for $\ga$ of
the order of several units. One can show that then the second term
in (\ref{jfp}) is dominant (the asymmetry of contributions from
the first and the second charges is due to our choice of the
reference frame: the first charge is at rest before the collision,
while the second is moving with the relativistic velocity). The
spectral distribution of radiation due to the second term extends
till $\om\sim 2\ga^2/d $ in the narrow angular region
$\delta\varphi\sim 1/\ga$ around the direction $\varphi=-\pi/2$,
while the first term has a lower ($\sim \ga/d$) frequency cutoff.
The total contribution of the second term to the radiation loss is
greater than that of the first one by a factor of $\ga$, so in
what follows we neglect the first term. Setting
$e_1=e_2=e,\,m_1=m_2=m$ we find from the Eq. (\ref{P1}) the
spectral-angular distribution of radiation as follows \be
\fr{dP^0}{d\om d\varphi}=\fr1{32\pi^2}\fr{e^6}{m^2}
\fr{\left[\cos^2\varphi+(\sin\varphi+\xi\ga^2)^2\right]\e^{-2\om
d\xi}}{\om\ga^6\xi^4}.\ee For $\ga\gg 1$, radiation  is peaked in
the direction of motion of the relativistic charge
$\varphi=-\pi/2$, and in the vicinity of this
direction~(cf.(\ref{xiultra})) \be\lb{xiultra1}\xi\thickapprox
\fr{1}{2}(\ga^{-2} +\beta^2), \ee where $\beta\ll 1$, and \be
\fr{dP^0}{d\om d\beta}=\fr{e^6\ga^2}{8\pi^2 m^2\om}
\fr{(1-\ga^2\beta^2)^2}{(1+\ga^2\beta^2)^4} \exp\left({-\fr{\om
d(1+\ga^2\beta^2)}{\ga^2}}\right).\ee The spectrum has an infrared
divergence, and in the forward direction $\beta=0$ it extends up
 to $\om\sim\om_{{\rm max}}$, where \be  \om_{{\rm
 max}}=\fr{\ga^2}{d}.\ee Integrating over frequencies from the
 infrared cutoff $\Delta^{-1}$ we obtain \be\lb{Pbeta}
\fr{dP^0}{d\beta}=
 \fr{e^6\ga^2}{8\pi^2 m^2 }
\fr{(1-\ga^2\beta^2)^2}{(1+\ga^2\beta^2)^4} {\rm  Ei} \left(1,
\fr{d(1+\beta^2\ga^2)}{\ga^2\Delta}\right).\ee Taking into account
relations (\ref{cors}) between parameters of 3+1 and 2+1 problems,
one can see that this expression coincides with (\ref{Pkapa}) for
$\ga=\kappa$, i.e. for $\al=0$:
\be\lb{Pbetas}\fr1{l}\fr{dP^0}{d\beta}=
 \fr{64 \pi^4 f^6\ga^2}{\mu^2 }
\fr{(1-\ga^2\beta^2)^2}{(1+\ga^2\beta^2)^4} {\rm  Ei} \left(1,
\fr{d(1+\beta^2\ga^2)}{\ga^2\Delta}\right).\ee
\section{Cosmological estimates}
In the standard axion cosmology scenario, which assumes the
Peccei-Quinn phase transition after inflation, the global  string
network is formed at the temperature $T_0\sim f$. Strings
initially move with substantial friction~\cite{DS87,VI91,MaSh97}
due to scattering of particles present in the hot cosmic plasma.
At some temperature $T_*<\f$ the  string network enters the
scale-invariant regime~\cite{HS91,Na97,YKY98} when strings move
almost freely with relativistic velocities~\cite{AlTu,Sh87} and
are able to form closed loops via interconnections. Finally, at
much lower temperature $T_1$ of the QCD phase transition, axions
become massive and strings disappear. Our mechanism presumably
should  work in the temperature interval $(T_*,\,T_1)$, though an
additional contribution from the damped epoch $T_0<T<T_*$ can also
be expected. The scaling density of strings is determined from
numerical experiments, it can be presented as \be\lb{ros}
\rho_s=\fr{\zeta\mu}{t^2},\ee with $\zeta$ varying from 1 to 13 in
different simulations.

Our aim is to estimate the total energy density $\varepsilon_a$ of
axions produced via the bremsstrahlung mechanism during the
cosmological period from $t_*$ (corresponding to temperature
$T_*$) till the QCD phase transition moment $t_1$. These are
commonly estimated as~\cite{BaSh97} \ba\lb{t*} t_* ({\rm sec})
&\sim & 10^{-20}\left(\fr{f}{10^{12}{\rm
GeV}}\right)^{-4},\\\lb{t1} t_1 ({\rm sec})&\sim & 2\cdot
10^{-7}\left(\fr{f}{10^{12}{\rm GeV}}\right)^{1/3}.\ea

Consider the scattering of an ensemble of randomly oriented
straight strings on a selected target string in the rest frame of
the latter. Since the dependence of the string bremsstrahlung on
the inclination angle $\al$ is smooth, we can use for a rough
estimate the particular result obtained  for the parallel strings
($\al=0$) introducing an effective fraction $\nu$ of ``almost''
parallel strings (roughly 1/3). Then, if there are $N$ strings in
the normalization cube $V=L^3$, we have to integrate the radiation
energy released in the collision with the impact parameter $d=x$
over the plane perpendicular to the target string using the
measure $N/L^2\cdot 2\pi x dx$. Actually we need the radiation
power per unit time, so, for an estimate, we have to divide the
integrand on the impact parameter. This quantity should be
multiplied by the total number of strings $N$ to get the radiation
energy released per unit time within the normalization volume.
Therefore, for the axion energy density generated per unit time we
obtain: \be\fr{d \var_a}{dt}=\int_0^L
P^0\,\nu\,\fr{N}{L^2}\,\fr{N}{V}\,2\pi dx.\ee As an expression for
$P^0$ we will use here the Eq. (\ref{final}), and we have to
integrate over the impact parameter. Taking into account that the
string number density is related to the energy density (\ref{ros})
via \be \fr{N}{V}=\fr{\rho_s}{\mu L},\ee introducing the integral
\ba \lb{y} \xi(w)&=&\int_0^w \phi(y) dy=12\sqrt{\fr{w}{\pi}}
\Biggl[ \ _2F_2\left(-\fr12,- \fr12; \fr12, \fr32;-w\right) -\nonumber\\
&-&{_2F_2}\left(- \fr12,- \fr12; \fr12, \fr12;-w\right)\Biggr]+
\left(\fr{13}{2}-3\ln\left(4 w\e^{C}\right)\right)w,\ea
and
using for $\mu$ the expression (\ref{mure}) with only the second
(leading) term, we obtain \be \fr{d\var_a}{dt}\sim K
f^2\ga^3\xi(w)
 \fr1{t^3\ln^2(tf)},\ee where \be
K=\fr83\pi^4\nu\zeta^2,\quad w=\fr{L}{\ga^2\Delta}.\ee Since in
the cosmological context $L\sim\Delta\sim t$, one can take
$w=\ga^{-2}$ for an estimate.

\begin{figure}
\centering
\includegraphics[width=6cm,height=6cm,angle=-90]{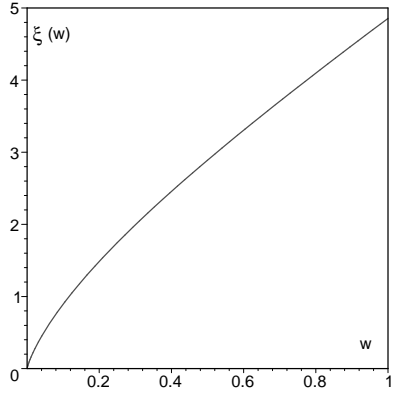}
\caption{Dependence of the axion radiation rate  on the Lorentz
factor of the collision $\xi(w),\,w=L/(\ga^2\Delta)$.}
\end{figure}
The realistic value of $\ga$ is of the order of unity, while our
formulas were obtained in the $\ga\gg 1$ approximation, since we
were keeping only the second term in the exact amplitude
(\ref{jfkappa}). But an independent calculation shows that for
$\ga\sim 1$ the contribution from the first term is of the same
order, so we can hope to get reasonable order of magnitude
estimate using the above formulas for $\ga$ not large.  The full
$\ga$-dependence is given by the function $\ga^3\xi(\ga^{-2})$,
where $\xi(w)$ is plotted on Fig. 5. For
$\ga=\sqrt{2}$\be\lb{1}\ga^3\xi(\ga^{-2})\approx 8.137,\ee while
for $\ga=5$\be\lb{2}\ga^3\xi(\ga^{-2})\approx 55.83.\ee Another
source of uncertainty is the numerical value of the parameter
$\zeta$, which was obtained in different simulations within the
region between 1 and 13. Now we have to integrate the radiation
power over time from $t_*$ to $t_1$ as given by (\ref{t*}) and
(\ref{t1}). Using the integral (with substitution $x=1/t$)
\be\int\limits\fr{x}{\ln^2x}dx=-\fr{x^2}{\ln x}+2{\rm Li}(x^2),
\ee where the logarithmic integral function is introduced, which
for large $x$ is equal approximately to \be -\fr{x^2}{\ln x}+2{\rm
Li}(x^2)\approx\fr{x^2}{2\ln^2x},\ee we find the energy density of
axions at the moment $t_1\gg t_*$: \be \var_a \sim \fr{K
f^2\ga^3\xi(w)}{2t_*^2\ln^2(t_*/t_0)},\ee where the ratio
$t_*/t_0$ can be estimated as \be
\frac{t_*}{t_0}=\left(\frac{m_{\rm Pl}}{f}\right)^2.\ee Finally,
dividing by the critical energy density at $t=t_1$
\be\varepsilon_{\rm cr}=\fr{3m_{\rm Pl}^2}{32\pi t_1},\quad m_{\rm
Pl}=1.22\cdot 10^{19}{\rm GeV},\ee we obtain the following
estimate for the relative contribution of bremsstrahlung axions
\be\Omega^{\rm br}_a\sim 0.5\cdot
10^{16}\left(\frac{\zeta}{13}\right)^2\left(\frac{f}{10^{12}{\rm
GeV}}\right)^{32/3},\ee where we have used the value (\ref{1}).
This estimate shows that our new mechanism is very sensitive to
the value of $f$ and gives an upper bound on axion window of the
order of $10^{10} {\rm GeV}$. We postpone more precise
cosmological estimates for a separate publication. In particular,
more careful analysis is needed to find radiation loss for
moderately relativistic string velocities.
\section{Conclusions}
In this paper, we have suggested a new mechanism for the axion
emission by the global string network: the bremsstrahlung under
string collisions. As far as we are aware, this effect was not
discussed in the context of the cosmic string scenario so far.
Though it is of the second order in the axion coupling constant,
rough cosmological estimates show that it is not small and gives a
contribution of the same order of magnitude as radiation due to
oscillating loops.

We have found the radiation amplitude using the perturbation
theory whose validity is restricted by relativistic (though not
necessarily ultrarelativistic) velocities of  colliding strings.
The frequency spectrum has an infrared divergence which is not an
artifact of our approximation, but rather is the effect of the
space-time dimensionality: as we have shown, an equivalent
description of the axion bremsstrahlung from strings in $3+1$
dimensions is provided by the electrodynamics of point charges in
$2+1$ dimensions, where its origin lies in the logarithmic
dependence on distance of the Coulomb potential. Unfortunately the
integration over the spectrum and the angular distribution can be
performed analytically only in the ultrarelativistic case, and the
final formulas obtained relate to this situation. But, in view of
a smooth dependence of the radiation efficiency on the Lorentz
factor of the collision, we hope that our results give a
reasonable estimate of the effect for values of the Lorentz factor
of the order of several units as well.

From  a purely theoretical viewpoint two aspects are worth to be
discussed. The first is the Cerenkov nature of radiation which is
associated with the possibility of the superluminal motion of an
effective source located around the point of minimal separation
between the strings. In perturbative setting, the strings, which
are straight in zero order approximation, get deformed under axion
interaction, with the deformation being maximal near this point.
This region therefore acts as an effective source of radiation
arising in the next order approximation. This explains why
radiation is concentrated on the Cerenkov cone directed along the
trajectory of the effective source. Another interesting feature is
that the effect has an alternative $2+1$ interpretation as the
bremsstrahlung of point charges. This is due to the fact that two
strings inclined with respect to each other and moving in the
superluminal regime can be made parallel by suitable
reparameterizations of their world-sheets and the choice of the
space-time Lorentz frame. This transformation is only feasible
once the relative string motion is superluminal in the sense
described above. Classical dynamics of parallel strings
interacting via the axion field in $3+1$ dimensions is equivalent
to dynamics of point charges in $2+1$ electrodynamics, therefore
the string axion bremsstrahlung is reduced to the point particle
bremsstrahlung in lower dimension.
\section*{Acknowledgments}
The final version of this work was completed while one of the
authors (DVG) was visiting Erwin Shr\"odinger Institute (Vienna)
under the programm "Gravity in two dimensions". He is grateful to
Organizers for the invitation and support and to participants of
the Workshop for stimulating discussions. The work of DVG was also
supported in parts by the Ministry of Education of Russia.

\end{document}